\documentclass[a4paper,twocolumn,amsmath,amssymb,showpacs,floatfix,prl,preprintnumbers,footinbib]{revtex4}
\usepackage{graphicx}
\usepackage{dcolumn}% Align table columns on decimal point
\usepackage{bm}% bold math

\usepackage{color}

\begin{document}

\title{Imaging the lateral shift of a quantum-point contact using scanning-gate microscopy}
\author{S.~Schnez\footnote{Author to whom correspondence should be addressed: schnez@phys.ethz.ch}, C.~R\"ossler, T.~Ihn, K.~Ensslin, C. Reichl, and W. Wegscheider}      

\affiliation{Solid State Physics Laboratory, ETH Z\"urich, 8093 Z\"urich, Switzerland}

\begin{abstract}
We perform scanning-gate microscopy on a quantum-point contact. It is defined in a high-mobility two-dimensional electron gas of an AlGaAs/GaAs heterostructure, giving rise to a weak disorder potential. The lever arm of the scanning tip is significantly smaller than that of the split gates defining the conducting channel of the quantum-point contact. We are able to observe that the conducting channel is shifted in real space when asymmetric gate voltages are applied. The observed shifts are consistent with transport data and numerical estimations.
\end{abstract}

\pacs{72.20.-i, 73., 07.79.-v}

\maketitle

%%%%

Quantum-point contacts (QPCs) are the building blocks of most electronic nanostructures. Despite their conceptual simplicity, they show intriguing properties like quantized conductance \cite{Wharam1988, Wees1988} and rich many-body physics \cite{Cronenwett2002, Miller2007, Dolev2008}. Local-probe techniques like scanning-gate microscopy (SGM) proved to be particularly powerful tools to extract a wealth of information about QPCs and two-dimensional electron gases (2DEGs) in general \cite{Topinka2000, Topinka2001, Jura2007, Jura2010, Aoki2005}.

With the advent of high-mobility samples and the discovery of very fragile quantum effects -- most notable the 5/2-fractional quantum Hall state -- it is desirable to gain a quantitative understanding of local potential fluctuations in the 2DEG and the behavior of quantum states, in particular edge channels, in a QPC. In fact, conventional transport experiments were performed in which asymmetric top-gate voltages were used to tune the potential landscape in the QPC channel \cite{Williamson1990}. Here, we use SGM to image and quantify the shift of the conducting QPC channel. This is an essential part in understanding the influence of top-gates on the underlying high-mobility 2DEG.

% sample description, conventional transport

\begin{figure}
  \centering
  \includegraphics[width=\linewidth]{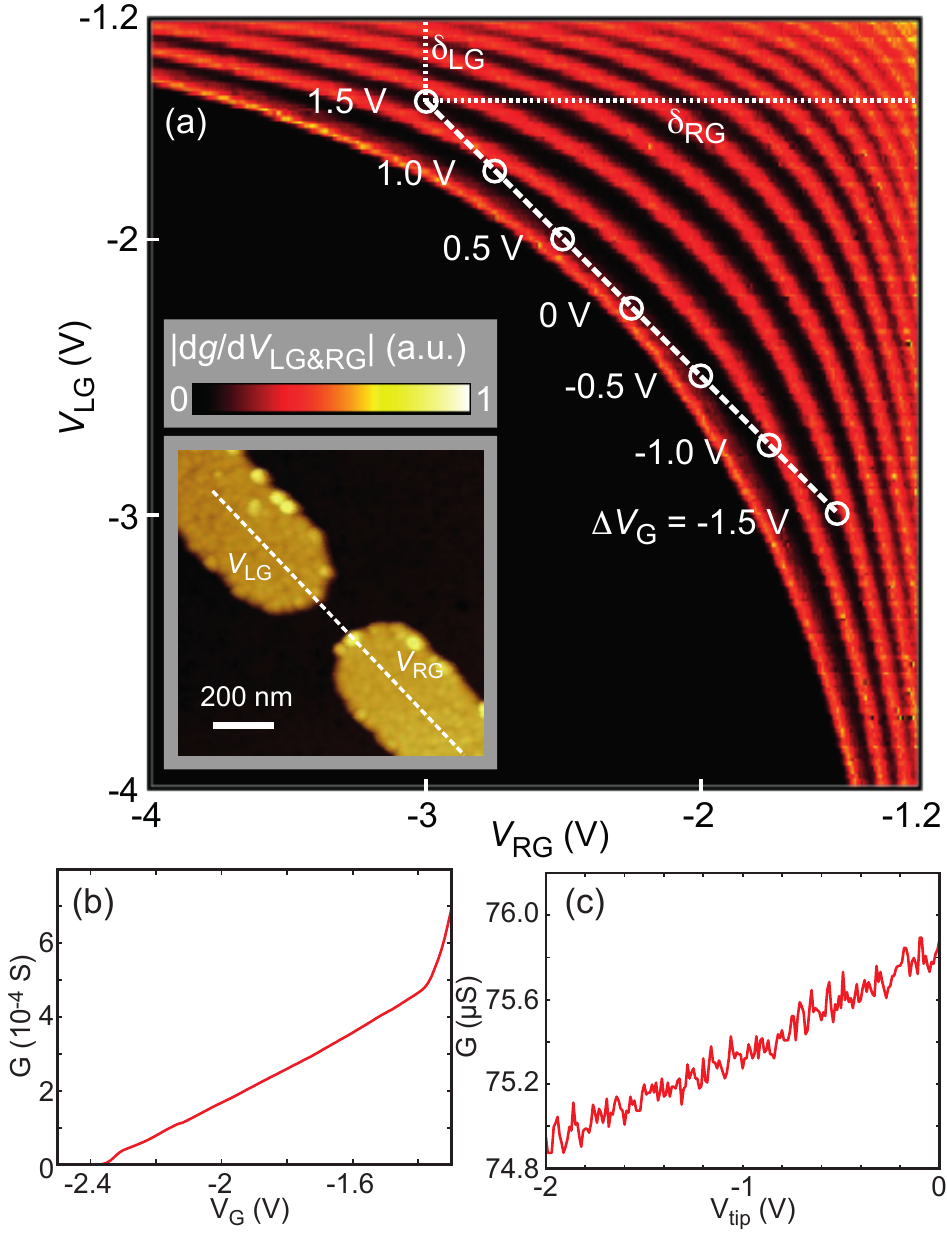}
  \caption{(Color) (a) Transconductance of the QPC from Ref.~\onlinecite{Roessler2011} at $T=1.3\,\rm{K}$. Up to 13 conductance steps appear as hyperbolic-shaped  features in the plane of the two top gates. We define the asymmetry in top gate voltages as  $\Delta V_{\rm{G}}:=V_{\rm{LG}}-V_{\rm{RG}}$. The white circles are points at the indicated values of $\Delta V_{\rm{G}}$. The quantities $\delta_{\rm{LG}}$ and $\delta_{\rm{RG}}$ are the distances of the conducting QPC channel to the left and right top gates as described in the main text below. Inset: Topography of a nominally identical QPC (discussed in Ref.~\onlinecite{Roessler2011}). The channel formed by the two top gates with applied voltages $V_{\rm{RG}}$ and $V_{\rm{LG}}$ has a width of 200 nm. (b) Depletion curve of the QPC under investigation here as a function of top-gate voltage. Depletion under the top gates is reached for $V_{\rm{G}} \leq -1.35\,\rm{V}$. Conductance steps are smeared out because of the rather high temperature of $T\approx 10\,\rm{K}$. Close to pinch-off, a weak 0.7-anomaly is observed. The QPC is pinched off at $V_{\rm{G}}\approx -2.35\,\rm{V}$. (c) Conductance of the QPC as a function of tip voltage $V_{\rm{tip}}$ for a tip height of 50 nm above the gates and $V_{\rm{G}}=-2.2\,\rm{V}$. \label{fig01}}
\end{figure}

The sample under investigation is a high-mobility GaAs-QPC with mobilities of more than $10\times 10^6\,\rm{cm}^2/\rm{Vs}$ at temperatures below 4.2 K. The measurements were performed at temperatures of around 10 K where mobilities are lower. The electron density is $n_{\rm{S}} = 3 \times 10^{15}\,\rm{m}^{-2}$ with a corresponding Fermi wavelength of $\lambda_{\rm{F}} = \sqrt{2\pi/n_{\rm{S}}}=44\,\rm{nm}$. %
The 2DEG resides in a quantum well (QW) which is buried 160 nm below the surface. In order to reach these high mobilities, the QW is symmetrically $\delta$-doped with Si-donors at depths of 70 nm and 250 nm. Remaining donor electrons which do not contribute to the 2DEG fill states in the $X$-valleys of AlAs layers close to the doping planes. Two further Si donor layers at 40 nm and 680 nm are expected to partially form DX-centers; their purpose is to compensate for surface states and substrate impurities. The gating properties of this kind of heterostructures are described in detail in Ref.~\onlinecite{Roessler2010}. It is not obvious {\it a priori} -- in particular because of hysteresis effects \cite{Roessler2010} -- whether SGM would work on such a device. 

The QPC is formed by two metallic top gates which form a gap of width $W=200\,\rm{nm}$ as shown in the inset of Fig.~\ref{fig01}(a). At low temperatures, conductance plateaus spaced by $2e^2/ h$ appear when the top gates are used to pinch off the conducting QPC channel. In Fig.~\ref{fig01}(a), such a pinch-off curve is shown for a device of (nominally) identical geometry \cite{Roessler2011}. The differential conductance $g = \textrm{d} I_{\rm{SD}}/\textrm{d}V_{\rm{SD}}$ is measured in four-terminal configuration via lock-in technique. After numerical derivation with respect to the gate voltages $V_{\rm{LG}}$ and $V_{\rm{RG}}$, the transconductance $\textrm{d}g/\textrm{d}V_{\rm{LG\&RG}}$ (along the bisecting line of the plane) is shown in color as a function of the voltages applied to the two top gates. The conductance steps show up as hyperbolicly shaped curves of elevated transconductance. We define the asymmetry of top-gate voltages as $\Delta V_{\rm{G}}:= V_{\rm{LG}}-V_{\rm{RG}}$. If we apply the same voltages to both top gates, we denote this with just $V_{\rm{G}}$. 

Conductance quantization is barely visible for the QPC under investigation here because of the elevated temperature, as seen in Fig.~\ref{fig01}(b). The 2DEG below the top gates is depleted at $V_{\rm{G}}\approx -1.35\,\rm{V}$ and the QPC channel is formed. It is pinched off at $V_{\rm{G}}\approx -2.35\,\rm{V}$. We note that these are about the same values as for the reference device shown in Fig.~\ref{fig01}(a) and discussed in Ref.~\onlinecite{Roessler2011}.

We place the metallic tip of our home-built  cryogenic atomic-force microscope 50 nm above the QPC channel (height above the metallic gates) and use it as a local gate by applying a voltage $V_{\rm{tip}}$ to the tip \cite{Ihn2004}. Negative tip voltages lead to a decrease in QPC conductance as expected and shown in Fig.~\ref{fig01}(c). Extrapolating the linear behavior of Fig.~\ref{fig01}(d) leads to an estimated pinch-off voltage of $V_{\rm{tip}}\approx -190\,\rm{V}$ which cannot be reached in our experiment. This is in contrast to the experiment performed by Topinka {\it et al.} \cite{Topinka2000, Topinka2001} which was interpreted in terms of backscattering of electrons off a tip-induced depletion pivot. We define the lever arms $\alpha_{\rm{G, tip}}=\textrm{d} G/\textrm{d} V_{\rm{G, tip}}$ of the top gates and the tip as the slope of the curves (in the linear regime) in Figs.~\ref{fig01} (b) and (c) and extract $\alpha_{\rm{G}}\approx 400 \,\mu\rm{S/V}$ and $\alpha_{\rm{tip}}\approx 400\,\rm{nS/V}$. The values for $\alpha_{\rm{tip}}$ vary by a factor of $\sim 2$ depending on the value of $V_{\rm{G}}$. The ratio $\alpha_{\rm{tip}}/\alpha_{\rm{G}}\approx 10^{-3}$ is very small. This may result from the particular sample design and the 2DEG being deep below the surface embedded between special doping layers.

% SGM

\begin{figure}
  \centering
  \includegraphics[width=  \linewidth]{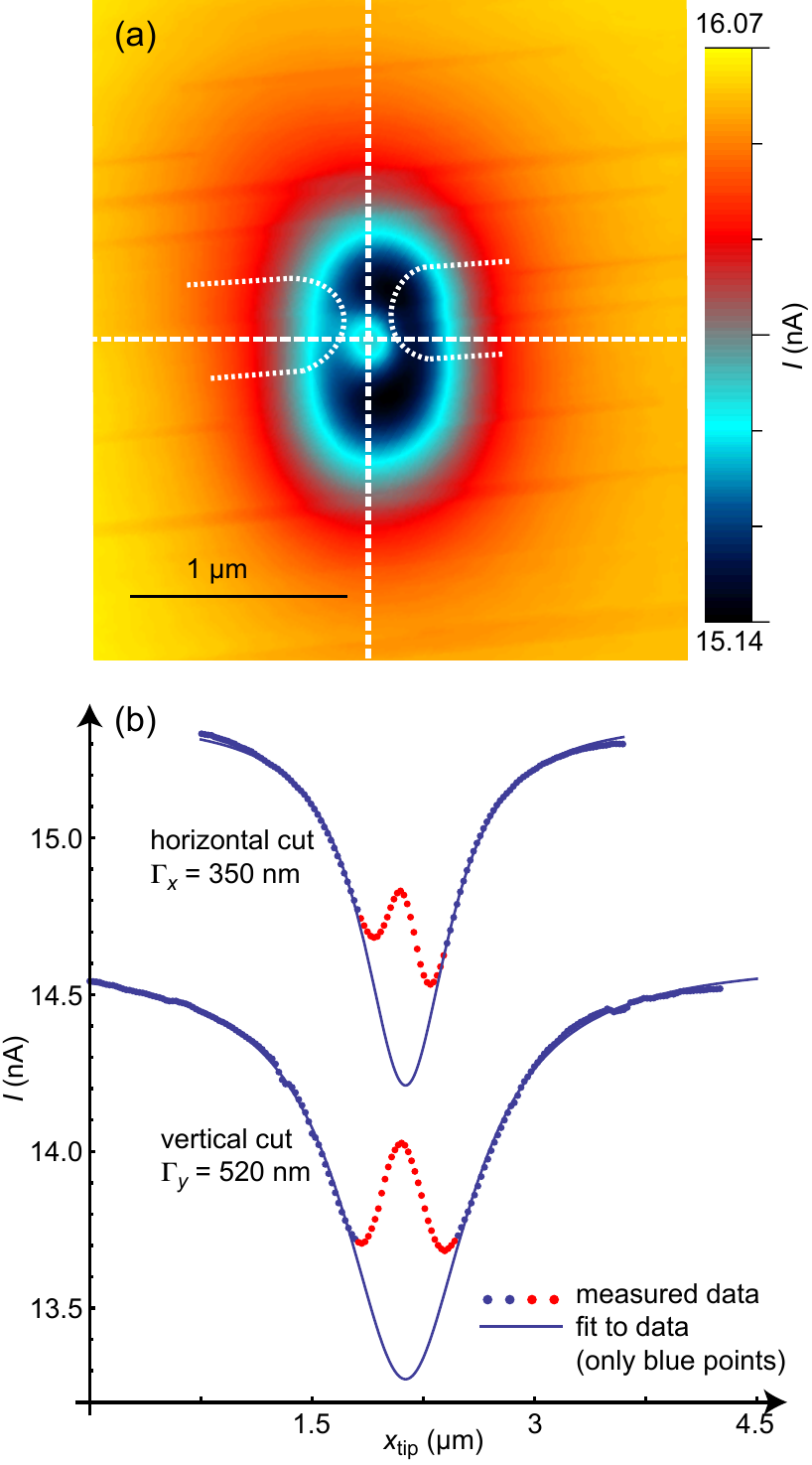}
  \caption{(Color) (a) Scanning-gate image of the QPC for a source-drain bias of $100\,\mu\rm{V}$, $V_{\rm{G}} = -2\,\rm{V}$, and $V_{\rm{tip}} = -3\,\rm{V}$. The current $I$ through the QPC is shown in color code as a function of tip position. The dotted lines trace the outline of the metallic top gates. (b) Line cuts along the vertical and the horizontal dashed lines in (a). They are offset from each other for clarity. The current  decreases when the tip approaches the QPC as it is expected for a negatively charged tip. The unexpected enhancement of current in the QPC shows up as a local maximum. The (blue) data points are fitted with a Lorentzian function (solid lines) with the indicated widths. \label{fig02}}
\end{figure}

We now perform SGM, meaning that we fix the top-gate and tip voltages and scan the tip at a constant height above the QPC. In this way, we obtain the spatially resolved current map shown in Fig.~\ref{fig02}(a). The dotted lines trace the edges of the top gates. The uncertainity of the alignment of the SGM image and the edges of the gates as determined from electrostatic-force microscoy corresponds roughly to the geometric width of the QPC channel.  Since the tip is negatively biased at $V_{\rm{tip}} = -3\,\rm{V}$ and the current $I$ is mainly determined by the conductance of the QPC, we expect the current to decrease when the tip approaches the QPC. The overall behavior does indeed follow the expectation. Unexpectedly, however, we find a spot of less suppressed current close to the center of the QPC where the two dashed lines cross in Fig.~\ref{fig02}(a) \footnote{This was confirmed using electrostatic-force microscopy obtained simultaneously with the SGM images (not shown here) so that other spurious effects can be excluded \cite{Schnez2011}.}. We will speculate on its physical origin below. The current map is asymmetric, i.e. lines of constant current have an elliptic shape. This might reflect the tip shape. However, since this asymmetry is aligned with the gate arrangement, we consider it more likely to be due to a screening effect of the tip-induced potential by the gates \cite{Schnez2011}.

Figure \ref{fig02}(b) presents line cuts along the horizontal and vertical dashed lines of the SGM image in Fig.~\ref{fig02}(a). The current decreases when the tip approaches the QPC until it increases again yielding a local current maximum which reflects the spot of less suppressed current of Fig.~\ref{fig02}(a). The width $\Gamma_{x,y}$ of the tip-induced potential is extracted by fitting the outer data points (blue data points without the enhanced current) with a Lorentzian curve of the form $I=I_0 \times \Gamma_{x,y}^2/\left(\left(x_{\rm{tip}}-x_{\rm{QPC}}\right)^2 + \Gamma_{x,y}^2\right)$, where we interpret $x_{\rm{QPC}}$ as the position of the QPC channel. The fits yield a width of $\Gamma_x = 350\,\rm{nm}$ and $\Gamma_y = 520\,\rm{nm}$ for the horizontal and vertical cuts, respectively. The smaller value of $\Gamma_x$ is expected because the tip-induced potential is wider than the QPC channel and is therefore affected by screening of the gates \cite{Schnez2011}.

% line sweeps

\begin{figure}
  \centering
  \includegraphics[width=  .93 \linewidth]{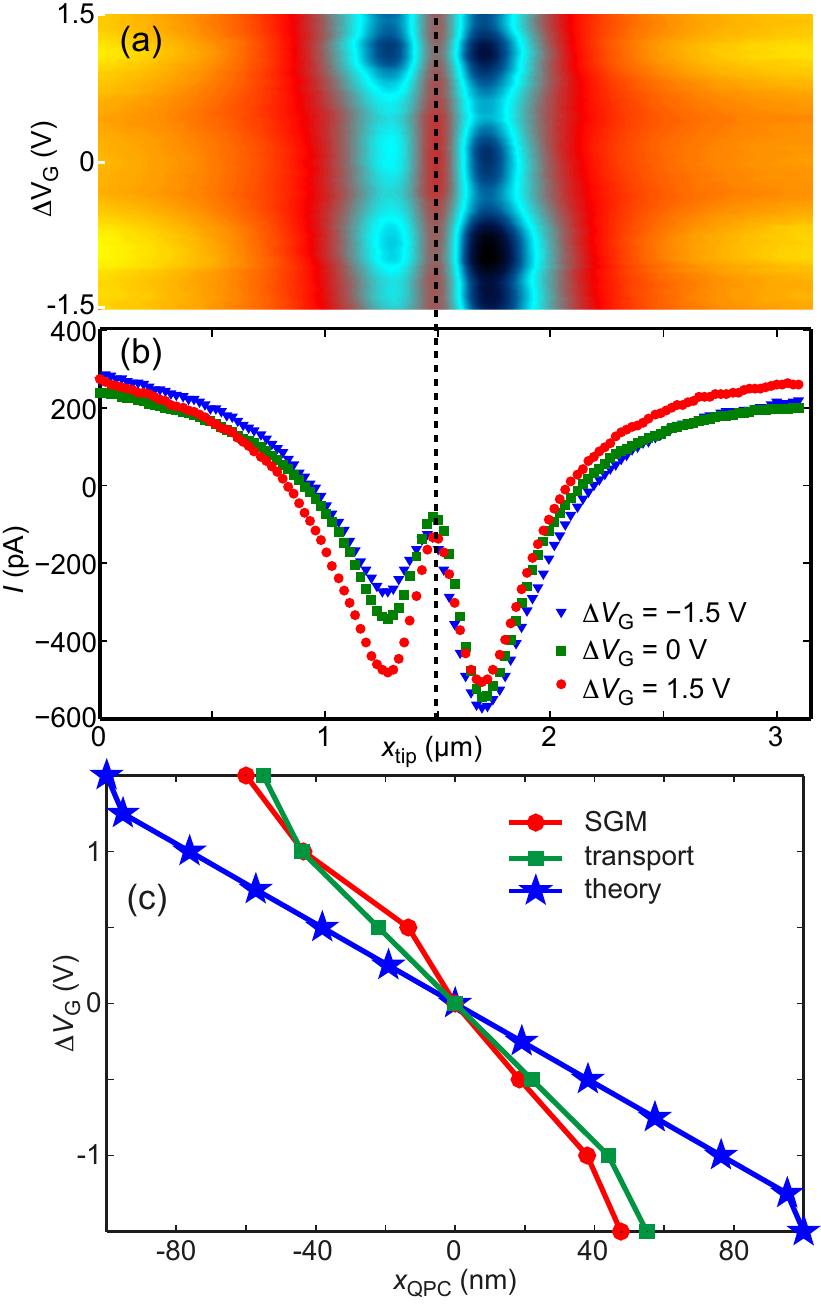}
  \caption{(Color) (a) Line scans of the tip along the QPC axis (dashed line in Fig.~\ref{fig01}(a), scale is given in panel (b), $V_{\rm{tip}} = -3\,\rm{V}$). The gate asymmetry is changed along the vertical axis such that $V_{\rm{LG}} = -2.25\,\rm{V} + \Delta V_{\rm{G}}/2$, and the current through the QPC is given in color code (arbitrary units, color palette as in Fig.~\ref{fig02}(a) but for each line the average value was subtracted). The position $x_{\rm{spot}}$ of the spot of enhanced current is independent of $\Delta V_{\rm{G}}$ as the vertical dashed line shows whereas the overall decrease in current is tilted to the left for more positive $\Delta V_{\rm{G}}$. This becomes clearer in panel (b) where we present line cuts extracted from the measurement shown in (a) for the indicated gate-voltage asymmetries: The peak of enhanced current is at the same tip position for all voltages but the outer envelope is shifted to the left for increasing asymmetry. (c) Shift of the QPC channel (red line/circles) for different gate-voltage asymmetries $\Delta V_{\rm{G}}$ deduced from panel (a) compared with the shift estimated from the transport experiment presented in Fig.~\ref{fig01}(b) (green squares) and with theory (blue stars).\label{fig03}}
\end{figure}

In order to further explore the properties of the QPC, we scan the tip along the horizontal dashed line in Fig.~\ref{fig02}(a) and change the asymmetry $\Delta V_{\rm{G}}$ according to $V_{\rm{LG}} = -2.25\,\rm{V} + \Delta V_{\rm{G}}/2$. The asymmetry then follows the dashed line in the plane of the two gate voltages in Fig.~\ref{fig01}(a). The result is shown in Fig.~\ref{fig03}(a), where the average current of each horizontal line was subtracted. Intuitively, one expects that the QPC channel is shifted laterally if the gate-voltage asymmetry is changed -- for more negative $\Delta V_{\rm{G}}$ the channel is pressed to the right and vice versa. Such a behavior was indeed theoretically predicted \cite{Glazman1991} and experimental results were interpreted assuming such a shift \cite{Williamson1990, Salis1999}. Nevertheless, a direct experimental observation of a shift has not been shown yet.

Our measurement shows, firstly, that the position of the spot of less suppressed current is observed at $x_{\rm{spot}} \approx 1.5\,\mu\textrm{m}$ (black, dashed line), independent of gate asymmetry (within the experimental accuracy). We therefore speculate that this spot is due to the geometrical arrangement of the gate electrodes. Otherwise, $x_{\rm{spot}}$ should depend on $\Delta V_{\rm{G}}$; in particular, $x_{\rm{spot}}$ should shift in parallel to $x_{\rm{QPC}}$ (see below) if the spot was due to a certain tip geometry. Secondly, the overall decrease in current -- measured by $x_{\rm{QPC}}$ as explained below -- is shifted to the left for more positive $\Delta V_{\rm{G}}$. This can be observed more clearly when looking at the line cuts for $\Delta V_{\rm{G}} = -1.5\,\rm{V}, 0\,\rm{V}, 1.5\,\rm{V}$ presented in panel (b): The local maxima of all curves indicating $x_{\rm{spot}}$ coincide with each other (vertical dashed line), but the overall decrease in current is shifted to the left for increasing asymmetry as intuitively expected. 

In Ref.~\onlinecite{Schnez2011}, a seeming shift of the position of a quantum dot was identified as being due to electrostatic effects; a real physical shift of the quantum dot could not be deduced. Here, we observe a physical shift of the QPC channel. If the shift was due to electrostatic effects, $x_{\rm{spot}}$ would show the same shift. Since this is not the case, the observed shift of the QPC channel is physical.

We follow the procedure described above for extracting the magnitude of the shift: Fitting the decrease in current with a Lorentzian yields $x_{\rm{QPC}}$ which we normalize such that $x_{\rm{QPC}}=0\,\rm{nm}$ for $\Delta V_{\rm{G}} =0\,\rm{V}$. In Fig.~\ref{fig03}(c), the position of the QPC channel extracted from Fig.~\ref{fig03}(a) for seven different gate-voltage asymmetries is plotted (red circles). The total shift is $\sim 110\,\rm{nm}$ in the investigated range of asymmetry.

The shift of the QPC channel can also be estimated from the transport data presented in Fig.~\ref{fig01}(a). The seven circles correspond to the gate-voltage asymmetries for which the shift was determined from the SGM data. The width $w$ of the QPC channel can be estimated from $w \approx n \lambda_{\rm{F}}/2$, where $n$ is the number of conducting QPC modes. The spatial distance $\delta_{\rm{RG}}$ from the channel center to the right top-gate edge is then estimated by half the Fermi wavelength times the number of modes that are added by following the dotted horizontal line until the 2DEG below the right top gate is not depleted anymore at $V_{\rm{RG}} = -1.2\,\rm{V}$. This yields the distance $\delta_{\rm{RG}} \approx 8 \lambda_{\rm{F}}/2 = 176\,\rm{nm}$. The corresponding analysis gives $\delta_{\rm{LG}} \approx 3\lambda_{\rm{F}}/2=66\,\rm{nm}$ for the distance from the channel center to the left top-gate edge. The sum $\delta_{\rm{RG}}+\delta_{\rm{LG}}$ is slightly larger than the separation $W=200\,\rm{nm}$ of the two gates. The procedure is repeated for the other six data points. The channel position is then defined as $x_{\rm{QPC}} := \left(\delta_{\rm{LG}} - \delta_{\rm{RG}}\right)/2$ and normalized as above. The positions of the QPC channel extracted this way are also plotted in Fig.~\ref{fig03}(c) as green squares. The agreement between both methods -- despite the rough estimate for the width of the QPC channel -- is very good and supports the assumption that a lateral shift of the QPC is both induced and detected in the presented experiment. This is the main result of this paper.

For completeness, we compare our experimental data with the analytic prediction for the total potential $\Phi (x,z)$ in the plane of the 2DEG from Ref.~\onlinecite{Glazman1991}. The $x$-axis is parallel to the dashed line in the inset of Fig.~\ref{fig01}(a); the $z$-axis is the normal to the surface of the semiconductor. The potential $\Phi$ is then the sum of two contributions of different origins,
\begin{equation}
	\begin{split}
		\Phi (x,z) = &\left( V_{\rm{LG}} \alpha_{\rm{LG}} \left(x,z, W\right) + V_{\rm{RG}} \alpha_{\rm{RG}} \left(x,z, W\right)\right)\\
		& + \Phi_{\rm{ion}}\left(x,z, W, n_{\rm{S}}\right),
	\end{split}
\end{equation}
where the first contribution depends on the gate voltages $V_{\rm{LG, RG}}$ with characteristic functions $\alpha_{\rm{LG, RG}}$ and the second is due to charged donors in the slit between the gates. This function has a minimum along the $x$-axis within the slit which forms the QPC channel. Its position $x_{\rm{QPC}}$ depends on the gate voltages and can thus be shifted as a function of asymmetry. We determine the minima for the parameters of our sample and the different gate asymmetries; the result is shown as blue stars in Fig.~\ref{fig03}(c). The kink at $\Delta V_{\rm{G}}=\pm 1.5\,\rm{V}$ occurs because in the model the QPC shift cannot exceed the width $W=200\,\rm{nm}$ of the QPC, i.e. $\left| x_{\rm{QPC}} \right| \leq 100\,\rm{nm}$. Theory and experiment agree within a factor of 2 which is reasonably good. The theoretical overestimation of the shift is due to the following two reasons: (i) In the theory, the length of the QPC channel is assumed to be much longer than its width $W$ and the depth of the 2DEG is assumed to be marginal compared to the width $W$. These conditions are experimentally not fulfilled. (ii) The X-electrons and the 2DEG itself can (self-consistently) screen the influence of the gates \cite{Laux1988} -- a possibility which is not implemented in the theory.

Finally, we want to comment on the physical origin of the spot of less suppressed current inside the QPC channel. The fact that its position $x_{\rm{spot}}$ is independent of applied gate voltages indicates that the origin is linked to the topographic gate-electrode arrangement. For example, the tip-induced potential may be screened efficiently by the gates at the spot such that the current through the QPC increases. An alternative, more subtle explanation is that the precise tuning of the QPC at the spot leads to a parallel conductance of $X$-electrons. Such an effect has been observed in conventional transport for certain gate voltages \cite{Roessler2011a}.

% conclusion
In conclusion, we presented scanning-gate measurements on a QPC processed on a high-mobility Al$_X$Ga$_{1-X}$As heterostructure \cite{Roessler2011}. Our experiments proved that it is possible to perform SGM on such a structure which is not obvious because of hysteretic effects \cite{Roessler2010}. However, it was not possible to deplete the 2DEG underneath the tip completely. This may open new possiblities to perform future experiments in the weakly invasive regime. In SGM images, an unexpected spot of less suppressed current showed up close to the center of the QPC. Its origin has not been completely understood yet but is probably due to a screening effect. Most importantly, we imaged directly how the QPC channel shifts when different asymmetric gate voltages were applied. A comparison with an estimate from direct transport data gives excellent agreement. The agreement with a theoretical model neglecting self-consistent effects is reasonable and deviations are qualitatively understood. For the future, SGM on high-mobility structures can be used to locally investigate electron-electron interactions \cite{Jura2010} -- this is of particular interest when these structures are tuned into the (fractional) quantum Hall regime. Our results presented here suggest that this endeavor should be feasible.

\section{Acknowledgement:} Financial support by ETH Z\"urich and the Swiss Science Foundation is gratefully acknowledged. Some figures were created using the WSxM-software \cite{Horcas2007}.

\end{document}